\documentclass[reprint,secnumarabic,amssymb, nobibnotes, aps, prd, superscriptaddress]{revtex4-2}

\usepackage[none]{hyphenat}
\usepackage{graphicx}
\usepackage{float}
\usepackage{xcolor}
\usepackage{scalerel}
\usepackage{mathtools} 
\usepackage{amsmath}
\usepackage{physics}
\usepackage[normalem]{ulem}
\definecolor{bluecitation}{RGB}{45, 48, 146}
\usepackage[colorlinks=true,allcolors=bluecitation]{hyperref}%

\begin{document}

%%%%%%%%%%%%%%%%%%%%%%%%%%%
%          TITLE
%%%%%%%%%%%%%%%%%%%%%%%%%%%

\title{Mapping the complete evolution of  magnetic excitation in beam-plasma system driven by an ultra-intense, femtosecond laser}%

%%%%%%%%%%%%%%%%%%%%%%%%%%%
% AUTHORS AND AFFILIATIONS
%%%%%%%%%%%%%%%%%%%%%%%%%%%

\author{Moniruzzaman Shaikh}%
\affiliation{Tata Institute of Fundamental Research, 1 Homi Bhabha Road, Mumbai 400005, India}

\author{Amit D Lad}
\affiliation{Tata Institute of Fundamental Research, 1 Homi Bhabha Road, Mumbai 400005, India}

\author{Devshree Mandal}%
\affiliation{Institute of Plasma Research, Gandhinagar, Bhat, Gandhinagar 382428 India}

\author{Kamalesh Jana}%
\affiliation{Tata Institute of Fundamental Research, 1 Homi Bhabha Road, Mumbai 400005, India}

\author{Deep Sarkar}%
\affiliation{Tata Institute of Fundamental Research, 1 Homi Bhabha Road, Mumbai 400005, India}

\author{Amita Das}%
\affiliation{Department of Physics, Indian Institute of Technology Delhi, Hauz Khas, New Delhi 110016, India}

\author{G Ravindra Kumar}%
\email{grk@tifr.res.in}
\affiliation{Tata Institute of Fundamental Research, 1 Homi Bhabha Road, Mumbai 400005, India}

\date{\today}%

%%%%%%%%%%%%%%%%%%%%%%%%%%%
%        ABSTRACT
%%%%%%%%%%%%%%%%%%%%%%%%%%%

\begin{abstract}

Plasmas are beset with instabilities of all types, hydrodynamic, magneto-hydrodynamic, and electromagnetic. These instabilities are complex, occur over a large range of temporal and spatial scales, are most often unmanageable, and have seriously challenged our efforts at applications, even as they have shed light on the understanding of the physics of plasmas in the laboratory and astrophysical environments. A major reason for our limited success in their containment is the lack of direct experimental information on their origins and evolution, both temporal and spatial. In plasmas produced by high-intensity, short, and ultrashort pulse lasers, our knowledge of the instability stems from the (secondary) signals they generate e.g. scattering of electromagnetic waves in the form of Raman or Brillouin scattering. Rarely, if ever, has a direct measurement been made of the instantaneous evolution of the instabilities in plasmas. In this paper, we present direct measurements of the femtosecond evolution of the electromagnetic beam-driven instability that arises from the interaction of forward and return currents in an ultrahigh-intensity laser-produced plasma on a solid target. We track the evolution from the initial linear stage to the later nonlinear stage, capturing the entire instability evolution- all in one experiment, via femtosecond pump-probe measurements of the giant (megagauss) magnetic field created during the intense laser-matter interaction.  Our experimental findings are excellently supported by particle–in–cell (PIC) simulations. These first results indicate heretofore unknown rich physics associated with beam-driven instabilities. In particular, it has been shown that the emission of electromagnetic radiation triggers the excitation of instability as negative energy modes feed from the free energy source associated with the beam propagation. This has been known in the context of gravitational interaction where the emission of gravitational radiation is responsible for driving certain negative energy modes in rotating black holes. In the electromagnetic context, this is the first such demonstration. Femtosecond laser-produced plasmas can serve as model templates for experimental instability studies across various branches of physics.

\end{abstract}

\maketitle

%%%%%%%%%%%%%%%%%%%%%%%%%%%
%      INTRODUCTION
%%%%%%%%%%%%%%%%%%%%%%%%%%%

\section{Introduction}\label{sec:intro}

The high energy density (HED) plasma created by ultrahigh intense, femtosecond laser pulses is a test bed for the fundamental understanding of intense light-matter interactions and a facilitator for tabletop creation of extreme states, shocks, and giant magnetic fields \cite{Kaw, AmitaRMPP}. They readily enable table-top simulations of astrophysics, novel particle acceleration schemes, ultra-bright X-ray sources, and light-driven nuclear physics. While these applications have aroused numerous, widespread investigations, several basic questions continue to challenge us. An important and ubiquitous one is about the onset and evolution of the instabilities in HED plasmas \cite{WeibelPRL1959}. In particular, the charged particle beam-plasma instabilities \cite{AmitaPRR2020} have recently become very important due to their role in the fast ignition of laser fusion and several astrophysical processes.

While there have been numerous simulation studies of this class of instabilities \cite{BretPRL2005, RemingtonScience1999, CalifanoPRE1997, CalifanoPRE1998}, experimental information is rare and highly wanted. More specifically, the dynamics, including the onset and early evolution of instabilities in HED plasmas have not been captured in any experimental study.

It is precisely this question that we address in this paper and present an important advance. In a single experimental setting, we study the origin, the linear growth phase, and the nonlinear stage of a crucial, widely prevalent beam-plasma instability that is very relevant to the transport of relativistic charged particle beams in supercritical, high-temperature plasmas, and the cause of giant magnetic fields in HED plasmas \cite{SandhuPRL2002, MondalPNAS2011, GChatterjeeNC2017, MShaikhPPCF2017} and in astrophysical systems. We accomplish this using a 25 femtosecond, multi-terawatt table-top laser, and pump-probe spatio-temporal polarimetry \cite{GChatterjeeRSI2014} of a HED plasma created on a solid target, achieving micron-scale spatial and femtosecond temporal resolution simultaneously. Our novel technique involved examining the evolving plasma at the plasma rear and the probe pulses that are reflected at the critical density surface which moves across the plasma on a femtosecond scale, as the plasma evolves, providing us an unprecedented opportunity to capture the beginning and linear stages of the instability. As earlier studies \cite{GChatterjeeNC2017} show, probing the plasma at the target front captures only the nonlinear stage. The examination of the plasma from the rear of the target (as shown in the schematic experimental set-up of FIG. 2) provides a view of the expanding plasma inside the target. This happens as the energetic electrons from the target front continuously dig into the target ionizing it to form plasma and the trailing electrons in the beam, in conjunction with electrons from the newly formed plasma, initiate beam plasma instabilities. The probe beam reaching the critical density layer from the rear thus always samples the newly formed plasma where the instability is initiated (illustrated in FIG. 2). Plasma density profile at two times $t_1$ and $t_2$ have been shown.  It can be observed that the probe pulse sent from the rear always samples the plasma activity in a freshly ionized layer.   On the other hand, the probe beam from the front observes a region of the plasma in which the instability has evolved since its inception and consequently is in the nonlinear phase. Thus the probes from the rear provide an unprecedented opportunity to look at the beginning (linear) stages of the instability. By employing a probe beam from the rear side of the target, we track the beginning, the linear growth stage, of the instability in a direct, unambiguous manner. Typically, probing the front surface reveals information about the nonlinear phase. The information about the nonlinear phase can also be extracted by probing the evolution of the magnetic field at the target rear when the electron beam ceases to ionize and form a fresh plasma layer and the critical density layer does not get pushed further at the rear.

In this paper, \textbf{Section II} explains a new experiment to study the time evolution of the beam plasma system, \textbf{Section III} displays the results of numerical simulations showing how the disturbances change over time while \textbf{Section IV} summarizes the findings of the study.

%%% Figure 1 %%%
\begin{figure}
    \includegraphics[width=1.0\columnwidth]{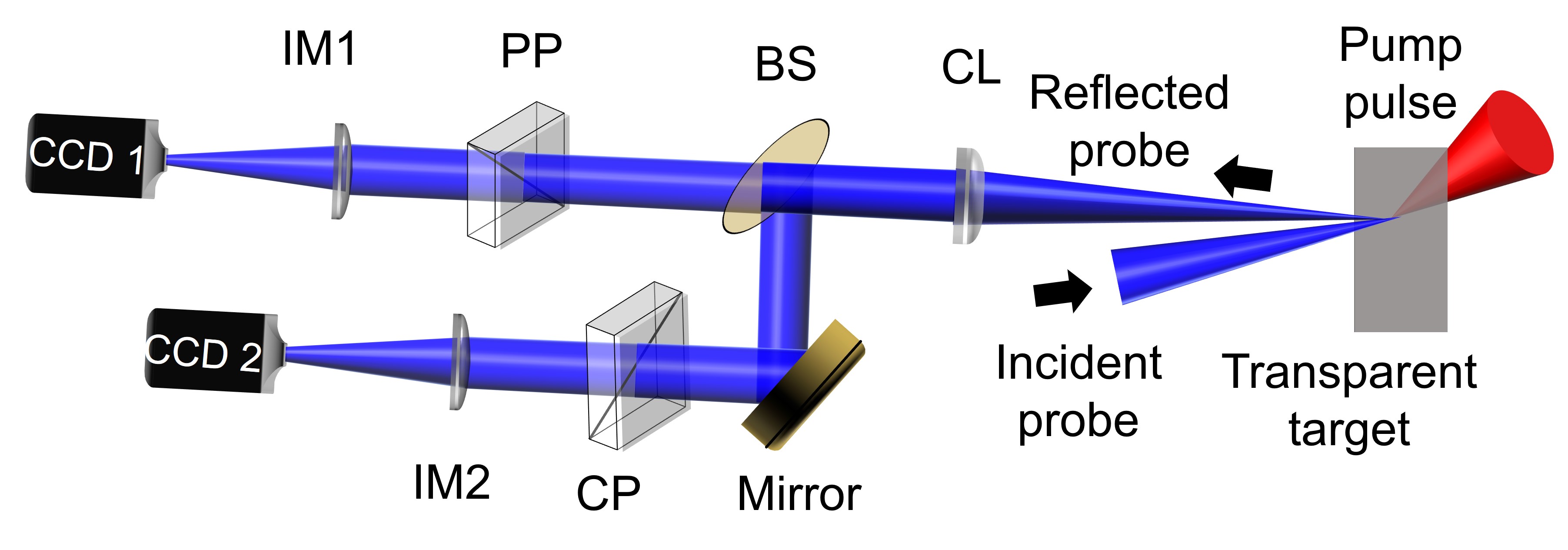}
    \caption{Pump-probe polarimetric setup to measure spatiotemporal evolution of ultra-intense laser-generated mega-gauss magnetic fields at the rear of thin targets. CL: Collimating lens, BS: Beam-splitter, PP: Parallel-polariser, CP: Cross-polariser, IM1, and IM2: Imaging lens 1 and 2, respectively.}
    %\label{fig:exp}
\end{figure}
%%% End Fig. 1 %%%

%%% Figure 2 %%%
\begin{figure}
\centering \includegraphics[width=\columnwidth]{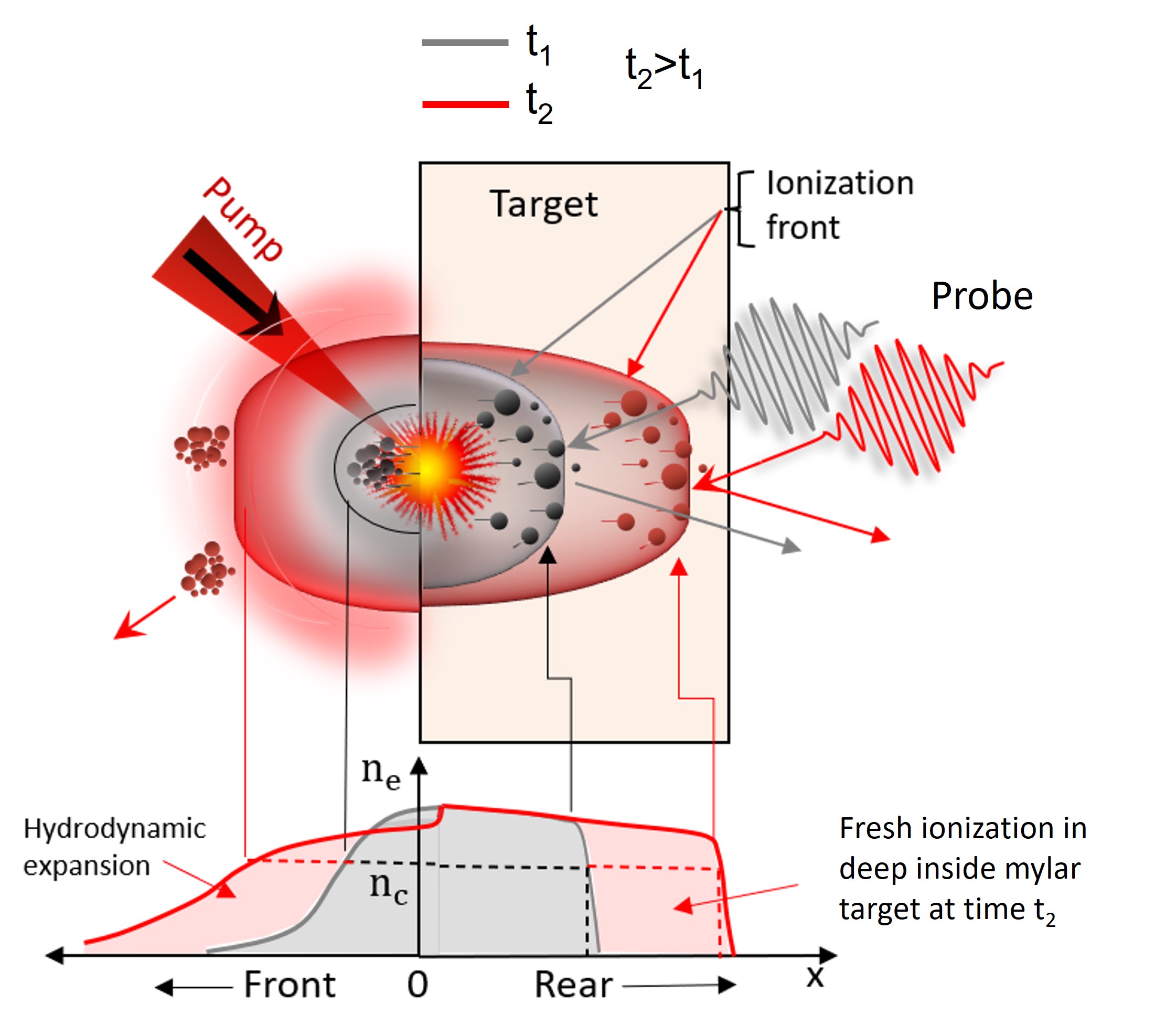}
    \caption{Schematic of the evolution of plasma density at various time delays, both in front of and deep inside the solid target. In front of the target, the probe pulse samples the plasma ahead of the critical surface, which has undergone hydrodynamic expansion. contrarily, at the rear side, the probe pulse consistently captures the freshly ionized plasma created by the propagation of the ionization front.}
    %\label{fig:LPI}
\end{figure}
%%% End Fig. 2 %%%

%%% Figure 3 %%%
\begin{figure}
\centering
\includegraphics[width=\columnwidth]{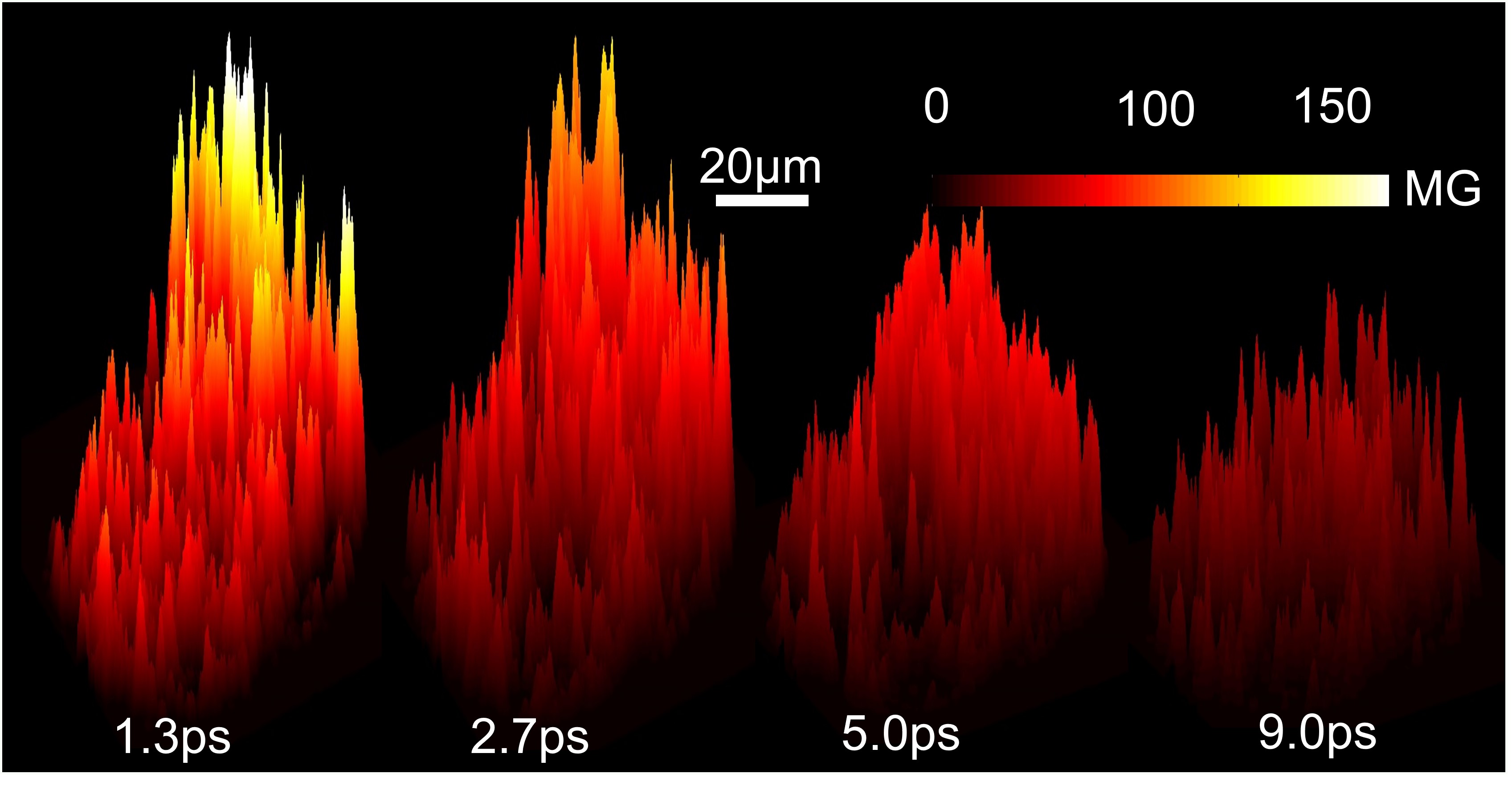}
    \caption{Experimentally measured magnetic field profile at the rear of 50 $\mu$m thick mylar target at a laser intensity of $3\times 10^{19}W/cm^2$.}
    %\label{fig:exp_2D_mylar}
\end{figure}
%%% End Fig. 3 %%%

%%% Figure 4 %%%
\begin{figure}
\centering
\includegraphics[width=\columnwidth]{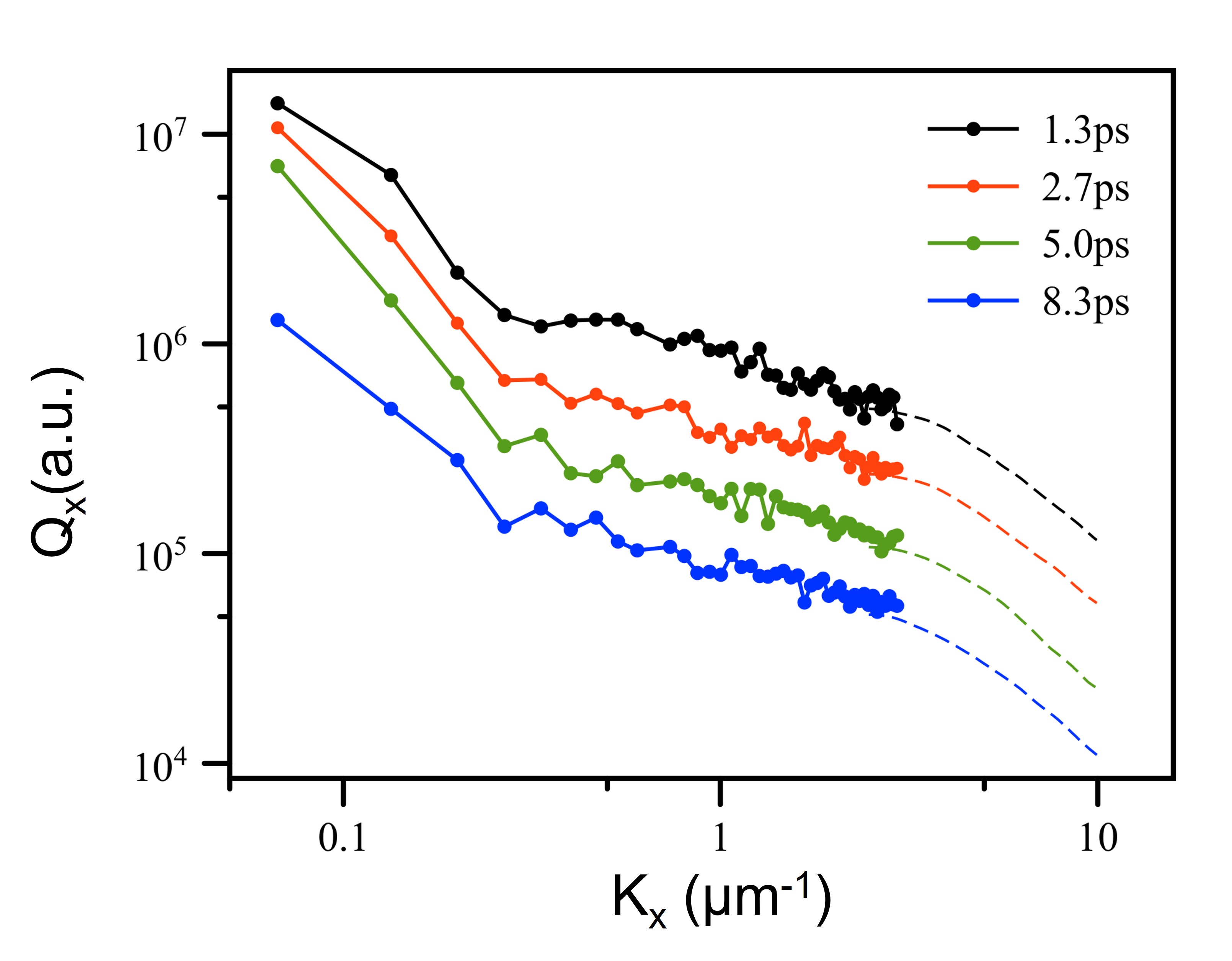}
    \caption{Experimentally measured power spectra at the rear of 50 $\mu$m thick mylar target at a laser intensity of $3\times 10^{19}W/cm^2$. The dashed lines indicate the trend.}
    %\label{fig:exp_theo_lineouts}
\end{figure}
%%% End Fig. 4%%%
%%%%%%%%%%%%%%%%%%%%%%%%%%%
%    EXPERIMENTAL SETUP
%%%%%%%%%%%%%%%%%%%%%%%%%%%

\section{Experimental setup}\label{sec:exp_setup}

The experiment was performed at the Tata Institute of Fundamental Research (TIFR, Mumbai) with a 100 TW laser system. The experimental setup is shown in FIG. 1. A p-polarized laser pulse (800 nm, $25$ femtosecond, and 1.0 Joule) was focused with an f/3 off-axis parabolic mirror to a 10 $\mu$m spot on the target at 45$^{o}$ angle of incidence, giving a laser intensity of $3\times 10^{19}W/cm^2$ on target. We used thin dielectric plastic (50 $\mu$m thick mylar) as targets. The high contrast of the laser pulse (picosecond pedestal/ peak intensity $10^{-9}$) ensures the creation of minimal pre-plasma during irradiation of thin targets. The target was translated by a computer-controlled X-Y stage to ensure a fresh spot during each laser shot. A small fraction ($5\%$) of the main interaction pulse was extracted with the help of a beam-splitter and up-converted to its second harmonic ($2 \omega$ = $400 nm$) using a $\beta$-barium borate (BBO) crystal, which was then used as the probe pulse. The $ 2\omega$ probe was appropriately attenuated and focused to a 75 $\mu$m diameter spot to image the rear of the plasma. The probe pulse was precisely delayed with respect to the main interaction laser pulse (precision of $7$ fs) using a computer-controlled delay stage. Note that the transverse spot of the probe pulse is more than seven times bigger than the transverse spot size of the pump pulse at the interaction region so that we could capture the magnetic fields at longer time delays when the plasma can undergo lateral expansion\cite{PKSinghPOP2013}. Additionally, we measured the space-integrated evolution of the magnetic fields by replacing the CCDs with two identical photo-diodes (PDs).
We captured the spatio-temporal dynamics of the giant magnetic fields generated during laser-matter interaction at relativistic laser intensity ($3\times 10^{19}W/cm^2$), where the probe laser pulse was sent at the target rear to image the magnetic field profiles. Probing the target rear with visible light (400 nm) is very different for transparent dielectrics (mylar) since the probe pulse can travel through the transparent target material till the critical density layer in the plasma. We relied on the Cotton-Mouton effect to measure the magnetic fields, where a normally incident, linearly polarized probe pulse gets elliptically polarized in the presence of giant magnetic fields as it transits the plasma and gets reflected at the critical surface for the 400nm probe. We measured the induced ellipticity for each pixel in the captured images of the probe pulse and retrieved the strength of the magnetic fields \cite{GChatterjeeRSI2014}. Figure 3 illustrates the transverse magnetic field profile captured from the rear plasma surface at various times.

To identify the prominence of the spatial scales that appear, we carry out the spatial Fourier transform and evaluate the power spectra of the magnetic field at various times. This has been shown in FIG.4. The power spectra are observed to be high at long scales right from the very beginning. At earlier times there is \textit{a clear dip} of spectral power at intermediate scales. This particular feature is in contrast with the data that has been observed when the front surface of the target is probed \cite{MondalPNAS2011, GChatterjeeNC2017, MShaikhPPCF2017} where the reflected probe beam samples the entire depth of the target, up to the critical layer inside the plasma. Thus, the spatial domain contains the integrated behavior of the nonlinearly perturbed medium. On the other hand, when the medium is probed from the rear, it only samples the freshly disturbed target region and shows the spectral behavior at a very early stage. Thus the difference between the spectral behavior when probed from the rear and front surfaces essentially mirrors the differences of excited magnetic scales at the early and later phases of evolution. We, therefore, infer that at the very early phase of evolution, the excitations at the skin depth scales are essentially absent and they appear only in later phases. It should be noted here that the typical well-known instabilities like Weibel \cite{WeibelPRL1959} and filamentation\cite{BretPRL2005, WhartonPRL1998} have a dominant growth rate at the skin depth scales. The absence of these scales at the earliest phase of evolution, then suggests that at the earliest time, scales are dominant for much longer than the skin depth scales. It is only later that the magnetic field power spectra as probed from the rear surface of the target show the appearance of the skin depth scales. In the next section, we try to understand this feature with the help of Particle-in-cell (PIC) simulations.

%%%%%%%%%%%%%%%%%%%%%%%%%%%
%        THEORY 
%%%%%%%%%%%%%%%%%%%%%%%%%%%

\section{Simulation}\label{sec:theory}

We have used the OSIRIS framework to carry out particle - in - cell (PIC) simulation \cite{HemkerArXiv2000}, \cite{osiris}, \cite{Fonseca2002} for studying the propagation of an energetic electron beam in the plasma medium. We have chosen a 2-D slab geometry with a $X-Y$ plane as a simulation domain. The simulation box has been chosen as  $L_x = 30L$ and  $L_y$ is $30 L $ in the simulations. Here $ L(=c/\omega_{pe})$ is the skin depth of the plasma. Here the beam propagation direction is along $y$. The plasma density $n_0 = 10^{21}  cm^{-3}$ is chosen to be homogeneous in the entire simulation domain. As an initial condition $10 \%$ of electrons within a transverse extent of $14L$ and longitudinal extent of $6L$ are given a directed velocity of $0.9c$ along $\hat{y}$. The simulation box is divided into $ 1500 \times 1500 $ cells with $ 8 \times 8$ particles per cell and the box has absorbing boundary conditions in both directions for particles and fields.

%%% Figure 5 %%%
\begin{figure}
\centering
\includegraphics[width=\columnwidth]{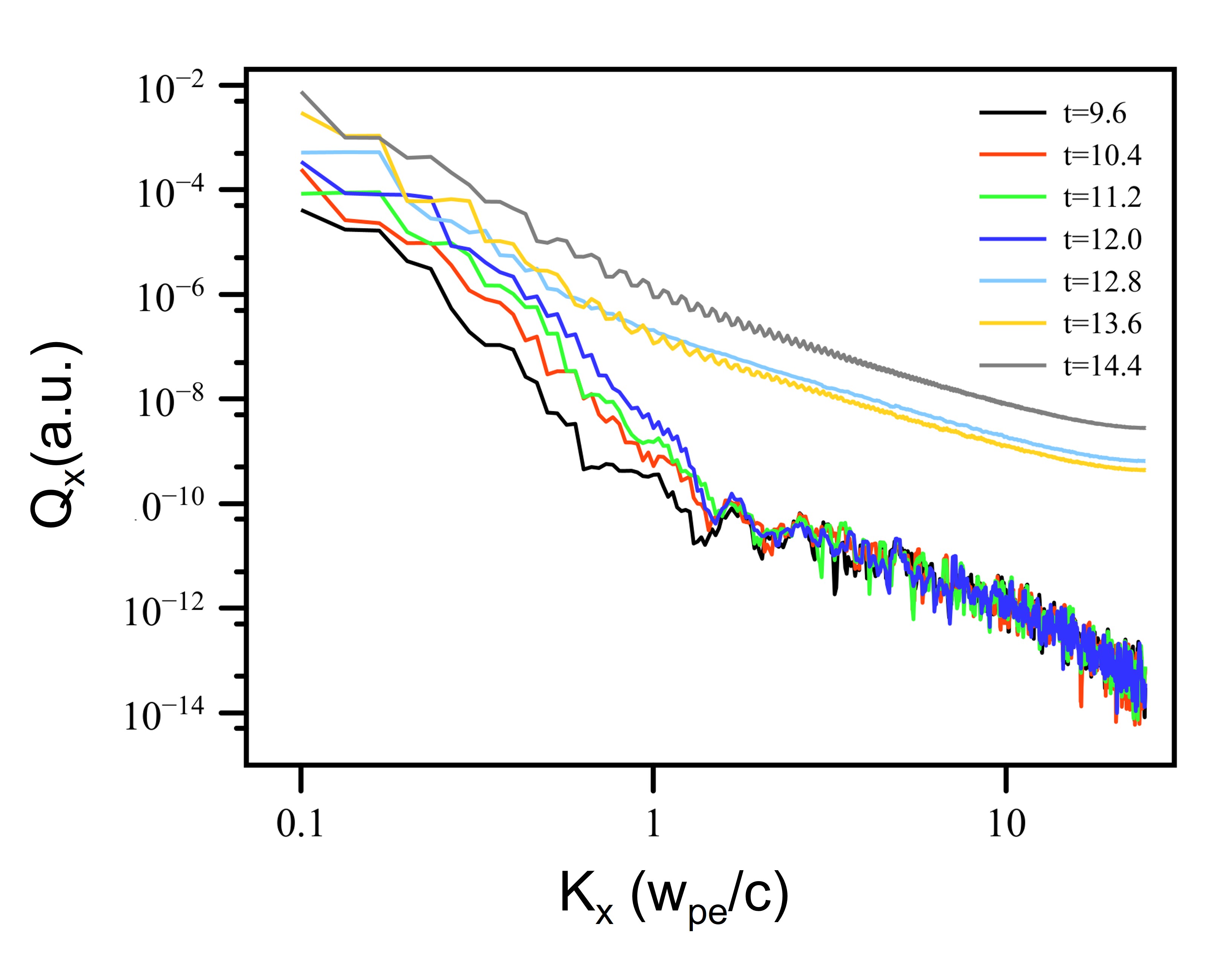}
    \caption{Simulated power spectra using $1/\omega_{pe}$ as the unit of time. The power spectra from t=9.6 to t=12.0 represent the linear phase of the instability, while t=12.8 to t=14.4 display the nonlinear growth of the beam-plasma instability.}
    %\label{fig:2DSimulationScheme}
\end{figure}
%%% End Fig. 5 %%%

%%% Figure 6 %%%
\begin{figure}
\centering
\includegraphics[width=\columnwidth]{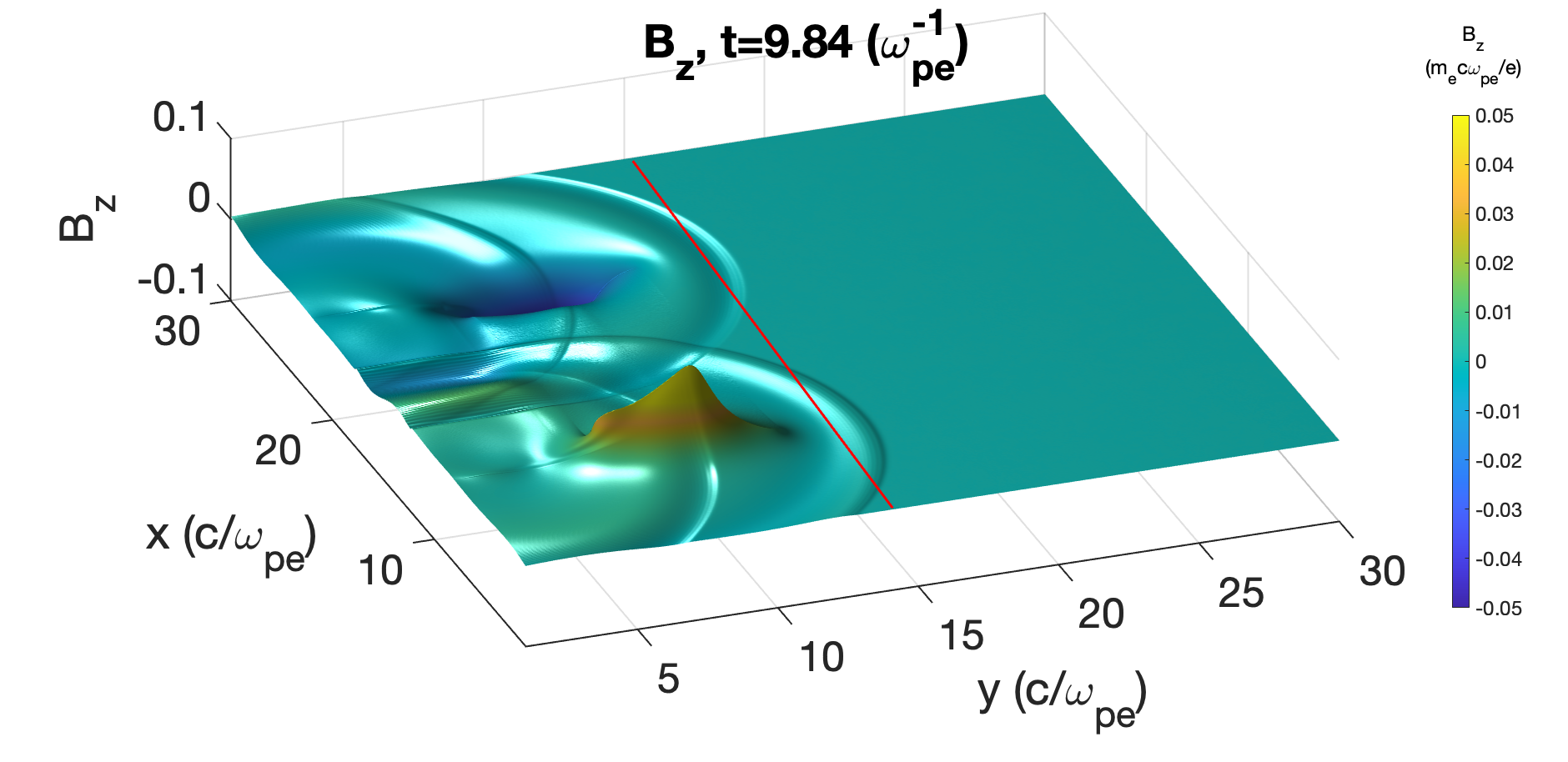}
    \caption{Simulated surface plot of the magnetic field. The linear region is on the right side of the red line.}
    %\label{fig:2DSimulationScheme}
\end{figure}
%%% End Fig. 6 %%%
 
As the beam propagates in the plasma it disturbs the plasma medium. The generation and evolution of the magnetic field around the tip of the propagating beam are of prime interest to us here, as this is the region that gets captured by the probe beam sent from the rear of the target in the experiment. The simulation provides for the spatio-temporal profiles of the magnetic field spectra in the 2-D plane. The surface plot of the magnetic field at $\omega_{pe} t = 9.84$ has been shown in Fig.6. The red line in the figure demarcates the region in the left and right domains. This line has been placed to demarcate the position towards its right being chosen around this time to capture the linear phase which subsequently provides the data also for the nonlinear phase as the beam moves with time.  The location of this particular line has been chosen by ensuring that the strength of the magnetic field in this region is typically two orders of magnitude smaller than the maximum magnetic field strength at this particular time. 
The Fourier transform of the magnetic disturbances as a function of $k_x$ over this limited region in $y$ (corresponding to the right of the demarcating red line) has been evaluated at various times. The spectra for this region have been obtained after integrating it over $dk_y$ to obtain power spectra as a function of $k_x$ in equation \ref{eq:1}.

\begin{equation}
Q_x = |B_k|^2(k_x) =   \int_{box}{|B_k|^2(k_x, k_y)dk_y}
\label{eq:1}
\end{equation}

The plot of $Q_x$ as a function of $k_x$  has been shown in FIG.5. At earlier times (till $t = 12.0$) the behavior of the spectra compares well with the experimental spectra shown in Fig.4 provided by the rear probe beam.  The experimental spectra shown in FIG.4 show a dip at central scales. This dip is also evident in the simulation observations.  It can be observed that subsequently, the simulation spectra show a characteristic change, and for time beyond $t = 12.8$, the dip disappears, and power law behavior indicates the onset of a nonlinear stage in this particular spatial domain is apparent. This characteristic is similar to the experimental data observed when the probe pulse is sent from the front and has been reported in earlier publications.

It is also interesting to note that in some other studies when energy is pumped at long scales a dip in spectra is observed. This happens as the nonlinear cascade rate is unable to push the energy pumped at long scales to smaller dissipative scales rapidly enough. As an example, the recent active fluid turbulence studies presented in FIG. 3 of Mukherjee \textit{et al.,} \cite{MukherjeeNatPhys2023}  also show such a dip at intermediate scales for large negative values of $\alpha$ (which implies energy injection at long scales). Magnetic field turbulence in a beam-driven plasma medium can also be viewed in a sense as active matter. The energy injection occurs due to inherent instabilities. In our case, the energy injection occurs due to a boundary-driven mechanism at long scales \cite{AmitaPRR2020}. Thus, until the nonlinear cascade rates remain small and have not disbursed energy to smaller scales and/or other instabilities (such as Weibel and Kelvin -Helmholtz etc.,)   which dump energy at skin depth scale do not appear, one observes a dip at the skin depth scale.

%%%%%%%%%%%%%%%%%%%%%%%%%%%
%   RESULTS and DISCUSSION 
%%%%%%%%%%%%%%%%%%%%%%%%%%%

\section{Results and Discussion}\label{sec:results}

We have observed both in experiments and simulations that the spectral power of the magnetic field associated with the initial excitations in the medium through the propagating electron beam has identical features. For both of them, (a) the power spectra are maximum at the most extended scales of the beam width dimension, (b) the spectra during the initial phase do not show a power law form, and (c) during the initial phase there is a dip in the spectra around electron skin depth scales. We now try to understand the physical reason for these characteristics of the spectra. The conventional beam plasma instabilities like filamentation, Weibel, and two-streams have the highest growth rate at the skin depth scale. Thus the earliest phase that has been captured in experiments with the probe beam entering from the rear side does not include the excitations of these modes. Yet the power level at much longer scales is present during this phase. This can only be understood provided there is a mechanism of long-scale magnetic field excitations present in the system. The possibility of such an excitation was discussed theoretically and demonstrated through simulations by Das \textit{et al.} \cite{AmitaPRR2020}. It was shown that the emission of electromagnetic waves from the boundaries results in the forming of long-scale (of the size of the beam width) magnetic field structures. These are essentially the negative energy modes excited by dissipation occurring from the emission of electromagnetic waves from the boundaries. The experiments here thus provide clear evidence of such a mechanism being operative in the context of laser-plasma experiments.

We would also like to point out that in the context of gravitational interactions, such a mechanism of gravitational wave emission leading to the excitation of negative energy modes from rotating black holes has already been described theoretically in detail by Chandrashekar \cite{ChandrasekharPRL1970, FriedmanAstro1978, BallaiAA2017}. We are demonstrating the presence of a similar mechanism for electromagnetic interaction. For the gravitational case, it is not possible to identify this mechanism with observational data as yet. We on the other hand are showing the presence of such a mechanism for electromagnetic interactions clearly through experiments and PIC simulations. Our experiments are thus a demonstration of yet another possibility wherein laser-plasma experiments transcend their limited scope to become relevant to other branches of science.
%%%%%%%%%%%%%%%%%%%%%%%%%%%
%    CONCLUDING REMARKS 
%%%%%%%%%%%%%%%%%%%%%%%%%%%

\section{Concluding remarks}\label{sec:conclusions}

The significant accomplishments of this study are as follows. It identifies novel diagnostics to fetch the spatial spectra of the magnetic field disturbances created by an electron beam plasma system during the earliest temporal phase of evolution.  The characteristic features of the magnetic field spectra are quite distinct from the power law behavior. This represents the early excitation of magnetic field disturbances which have not yet entered any nonlinear interaction regime. It is noteworthy that even though the magnetic field has significant power at long beam width scales, there is a clear absence of spectral disturbances around skin depth scales. This demonstrates that the conventional beam plasma instabilities are operative only at later times. At the beginning of time scales, some other mechanisms that generate scales much greater than the skin depth are present. Earlier \cite{AmitaPRR2020}, a mechanism was identified for the generation of these long-scale magnetic field structures. Its presence in experimental data was shown from the probe diagnostics involving the front surface. However, its appearance before the conventional instabilities was demonstrated \cite{AmitaPRR2020}  only through PIC simulations. Here, on the other hand, the onset of such a mechanism at the very beginning has been identified clearly through experiments. This reinforces our inference about the existence of such a mechanism and its appearance much ahead of the conventional modes often discussed in the context of beam plasma interactions.

It was identified by Das \textit{et al} \cite{AmitaPRR2020}, that the radiative emission of electromagnetic waves acts as a dissipation to drive the negative energy modes forming long-scale magnetic structure from the beam energy. In the context of rotating black holes it has been shown by Chandrashekar and others \cite{ChandrasekharPRL1970, FriedmanAstro1978, BallaiAA2017} that the loss of energy through gravitational radiation acts as a dissipation and leads to the excitation of negative energy modes. Our work, therefore, shows for the first time the electromagnetic counterpart of radiative dissipation leading to the excitation of negative energy modes.

%%%%%%%%%%%%%%%%%%%%%%%%%%%
%     ACKNOWLEDGEMENTS 
%%%%%%%%%%%%%%%%%%%%%%%%%%%

\begin{acknowledgments}
G.R.K. acknowledges partial support from J. C. Bose Fellowship Grant No.JCB-37/2010 of the Department of Science and Technology, Ministry of Science and Technology, Government of India. A.D. acknowledges support from the Science and Engineering Research Board (SERB), Department of Science and Technology (DST) Government of India, and the J. C. Bose Fellowship grant JCB/2017/000055 and CRG proposals CRG/2018/000624 and CRG/2022/002782.  A.D.L. acknowledges partial support from the Infosys-TIFR Leading Edge Research Grant (Cycle 2). A.D and D.M would like to acknowledge the OSIRIS Consortium, consisting of UCLA and IST (Lisbon, Portugal) for providing access to the OSIRIS  framework which is the work supported by NSF ACI-1339893. AD also acknowledges help from Laxman Prasad Goswami for the re-affirmation of simulation using various platforms. We thank Prof. K. Subramanian (IUCAA, Pune) and Prof. Dipankar Bhattacharya (RRI, Bangalore) for insightful discussions on the negative energy modes from black holes. We thank  Ankit Dulat for his help with the manuscript.
\end{acknowledgments}

%%%%%%%%%%%%%%%%%%%%%%%%%%%
%      BIBLIOGRAPHY
%%%%%%%%%%%%%%%%%%%%%%%%%%%

\bibliography{references}

\end{document}